\documentclass[conference]{IEEEtran}
\IEEEoverridecommandlockouts
\usepackage{cite}
\usepackage{amsmath,amssymb,amsfonts}
\usepackage{algorithmic}
\usepackage{graphicx}
\usepackage{textcomp}
\usepackage{xcolor}
\usepackage{algorithm}
\usepackage{subfigure}
\def\BibTeX{{\rm B\kern-.05em{\sc i\kern-.025em b}\kern-.08em
    T\kern-.1667em\lower.7ex\hbox{E}\kern-.125emX}}
\begin{document}

\title{Spectrum Focused Frequency Adversarial Attacks \\for Automatic Modulation Classification\\

}


\author{\IEEEauthorblockN{
\textsuperscript{1}Sicheng Zhang,
\textsuperscript{1}Jiarun Yu,
\textsuperscript{1}Zhida Bao,
\textsuperscript{2}Shiwen Mao,
\textsuperscript{1*}Yun Lin}

\IEEEauthorblockA{\textsuperscript{1}College of Information and Communication Engineering, Harbin Engineering University, Harbin, P. R. China \\
\textsuperscript{2}Department of Electrical \& Computer Engineering, Auburn University, Auburn, AL, USA \\
\textsuperscript{*}Corresponding Author:
linyun@hrbeu.edu.cn
 \\
}
}

\maketitle

\begin{abstract}
Artificial intelligence (AI) technology has provided a potential solution for automatic modulation recognition (AMC).
Unfortunately, AI-based AMC models are vulnerable to adversarial examples, which seriously threatens the efficient, secure and trusted application of AI in AMC.
This issue has attracted the attention of researchers.
Various studies on adversarial attacks and defenses evolve in a spiral.
However, the existing adversarial attack methods are all designed in the time domain.
They introduce more high-frequency components in the frequency domain, due to abrupt updates in the time domain.
For this issue, from the perspective of frequency domain, we propose a spectrum focused frequency adversarial attacks (SFFAA) for AMC model, and further draw on the idea of meta-learning, propose a Meta-SFFAA algorithm to improve the transferability in the black-box attacks.
Extensive experiments, qualitative and quantitative metrics demonstrate that the proposed algorithm can concentrate the adversarial energy on the spectrum where the signal is located, significantly improve the adversarial attack performance while maintaining the concealment in the frequency domain.
\end{abstract}


\section{Introduction}
With the continuous improvement of the level of informatization, national security, economic development, production and life are increasingly dependent on electromagnetic space.
The importance of awareness and regulation of electromagnetic space has been elevated to an unprecedented level.
Automatic modulation classification (AMC) has become a key technology in many applications such as cognitive radio, spectrum sensing, and spectrum management.
At the juncture where traditional methods show limitations, artificial intelligence (AI) technology opens new doors for researchers by virtue of low prior, strong fitting, and fast decision-making~\cite{9454255}.
Since then, methods suitable for various scenarios have emerged in an endless stream, continuously pushing up the performance~\cite{10.1007/978-3-319-44188-7_16,9200788,9392373,SichengZHANG12}.
However, the researchers found that adding carefully crafted and invisible perturbations to the input data can easily fool the model, which pose a great threat to the application of AI~\cite{2013arXiv1312.6199S}.
Undoubtedly, the study of electromagnetic space adversarial attack and defense is particularly critical for efficient, safe, and credible AI applications in electromagnetic space.

Once the electromagnetic space adversarial examples were discovered, it attracted the attention of researchers~\cite{2020arXiv201214392A}.
Sadeghi et al. firstly found that adversarial examples can seriously damage the performance of AI-based AMC models at a very small cost~\cite{8449065}.
In~\cite{9155389}, the aggressiveness of various gradient-based attack methods on the AMC model is studied, and the inverse relationship between the confidence of the signal category and the attack success rate is verified.
In~\cite{9259112}, the author carried out different adversarial attack methods against the AMC model, and studied the sensitivity of different types of data.
In~\cite{9609969}, the author proposes channel-robust channel reverse attack, and develops minimum mean square error attack and maximum received perturbation power attack.
In~\cite{9201397}, the authors consider the constraints such as the bit error rate, and generates signal adversarial examples that make the eavesdropper unable to identify correctly, but ensure that the cooperative receiver can receive it correctly.
In~\cite{9746669}, the authors consider the effect of random delay on the superposition of perturbations to the signal position, and propose a position-invariant adversarial attack method to effectively resist the impact of delay on attack performance.
In~\cite{9570781}, the authors launch an adversarial attack on the deep complex network model and use the logits combined evaluation metric to quantify the classification effect of the attacked model.

\begin{figure}[tbp]
\centering
\scriptsize
\subfigure[\scriptsize Original signal]{\includegraphics[width=0.48\linewidth]{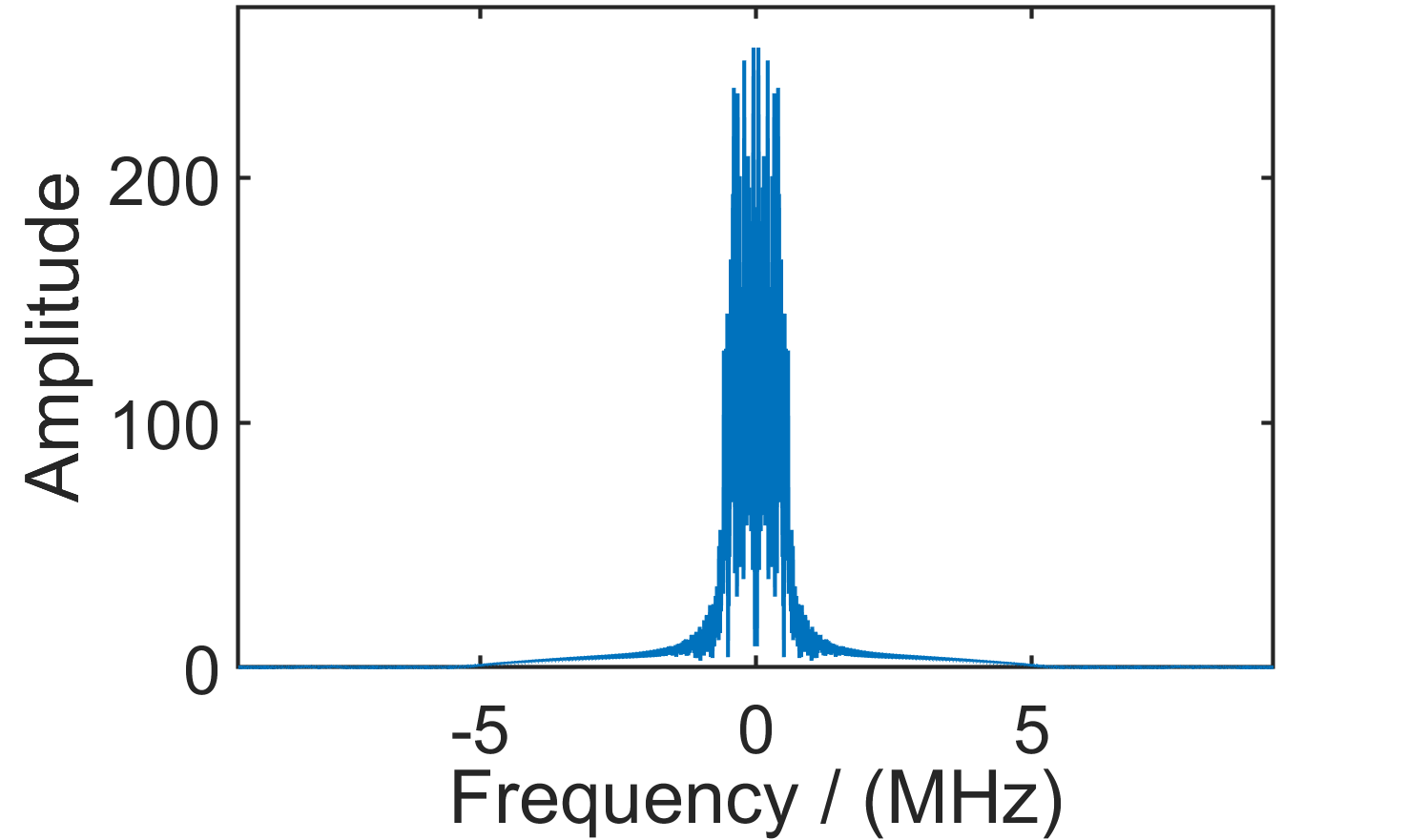}
\label{specOrg}}
\hfill
\subfigure[\scriptsize FGSM]{\includegraphics[width=0.48\linewidth]{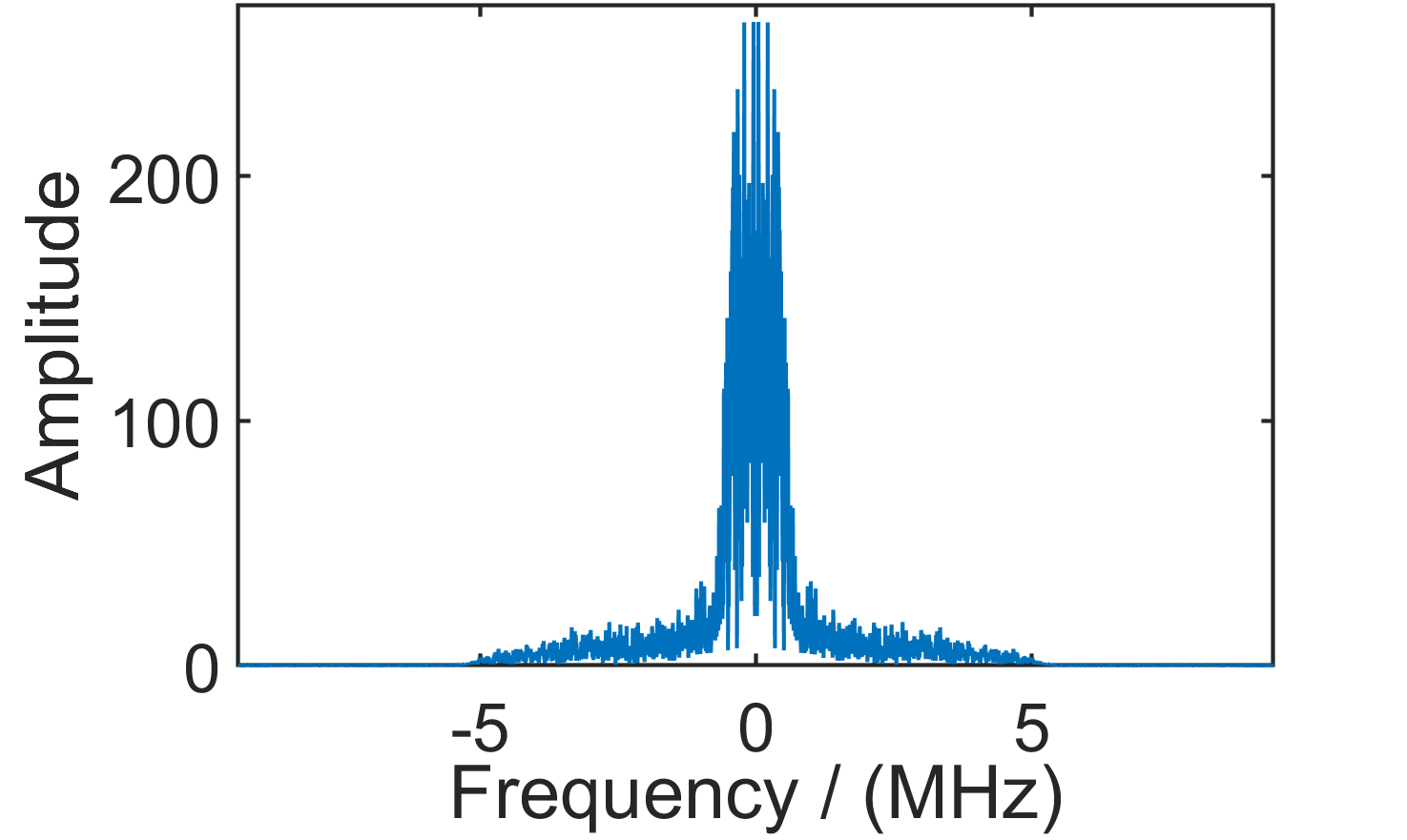}
\label{specFGSM}}
\hfill
\subfigure[\scriptsize PGD]{\includegraphics[width=0.48\linewidth]{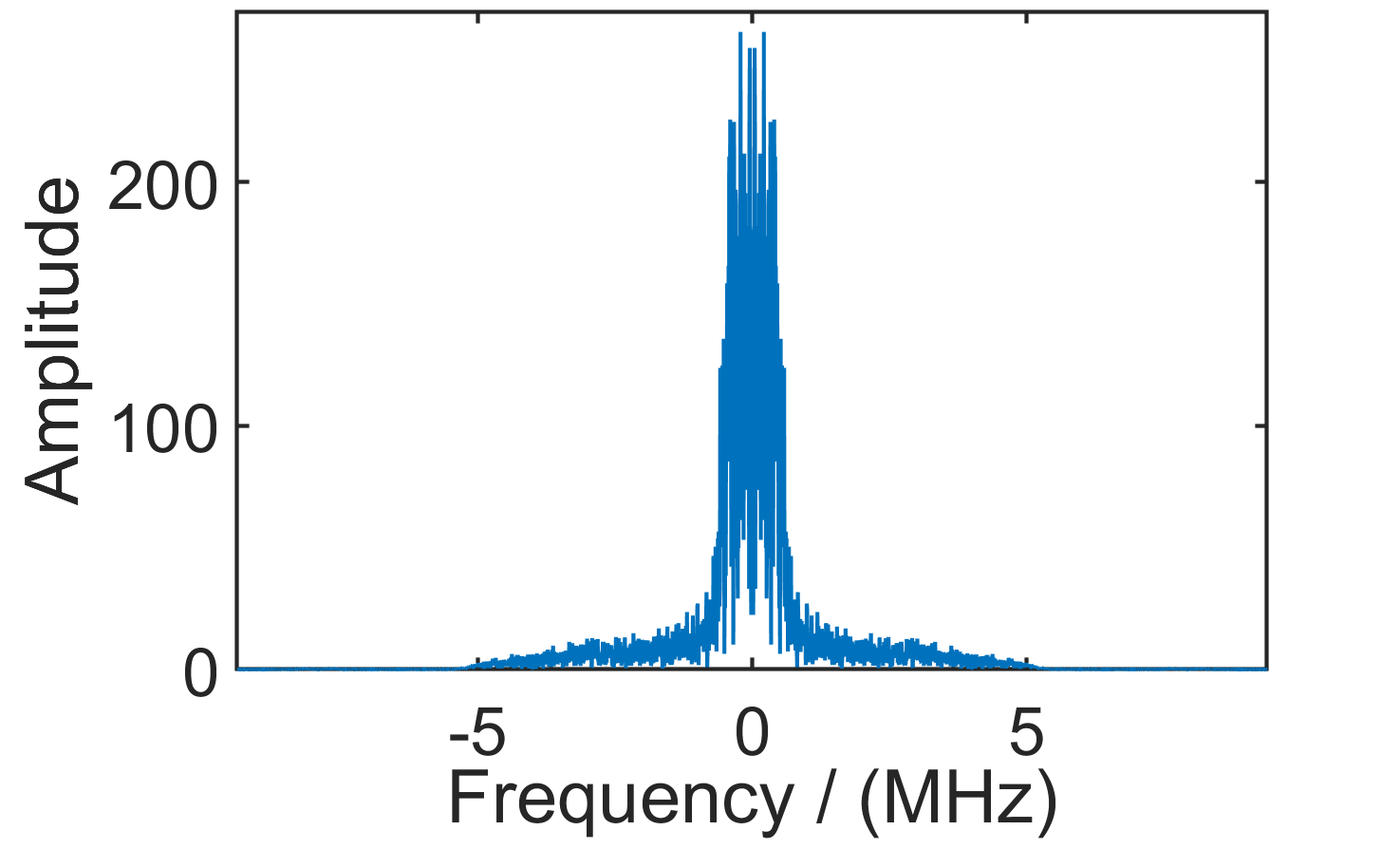}
\label{specPGD}}
\hfill
\subfigure[\scriptsize UAP]{\includegraphics[width=0.48\linewidth]{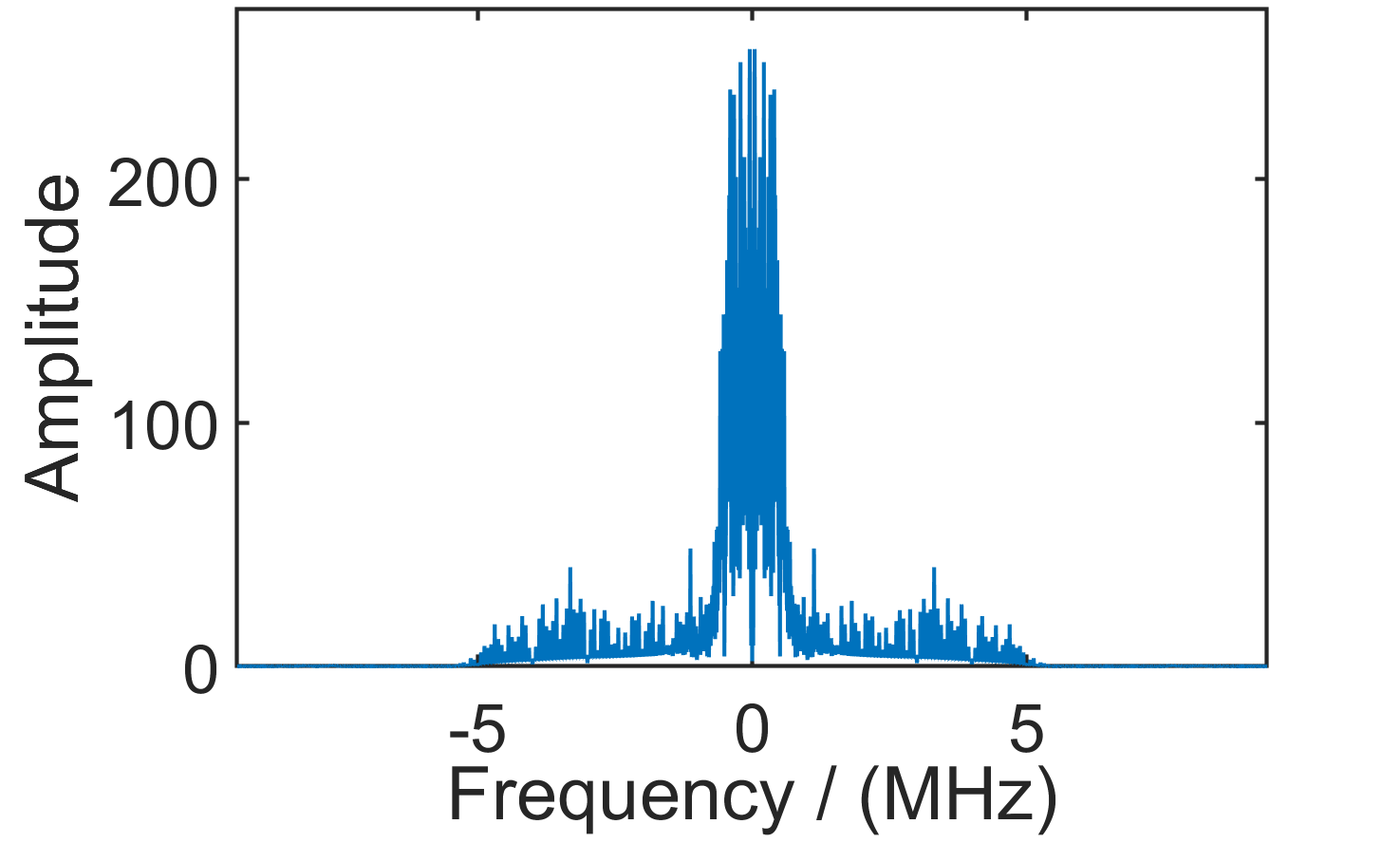}
\label{specUAP}}
\caption{Display diagram in frequency domain of existing attack methods.}
\label{specProblem}
\end{figure}

However, the existing adversarial attack methods are all designed in the time domain, using gradients to update the input data in a single step or iteratively with a certain step size.
The adversarial examples generated in this way will introduce more high-frequency components due to the abrupt changes in the time domain.
Compared to the spectrum of the original signal, the adversarial signal has obvious high frequency components in the sidebands, as shown in Fig.~\ref{specProblem}.
This will undoubtedly make the it easy to be detected in the frequency domain.
For this issue, we design the spectrum focused frequency adversarial attacks (SFFAA) algorithm from the perspective of frequency domain.
The algorithm can concentrate the adversarial energy on the spectrum where the signal is located.
In addition, we draw on the idea of meta-learning and design the Meta-SFFAA algorithm to improve the transferability in black-box attacks.
Extensive experiments, qualitative and quantitative metrics demonstrate that the algorithm can greatly improve attack performance while maintaining spectral concealment.

The rest of the paper is organized as follows.
Section \ref{sTwo} provides the system model and evaluation metrics.
Section \ref{sThree} describes the detailed pipeline of the proposed SFFAA algorithm, and Meta-SFFAA to in black-box scenarios.
Section \ref{sFour} introduces extensive experiments, conducts a comprehensive analysis of the proposed algorithm both qualitatively and quantitatively.
Section \ref{sFive} concludes the paper.

\section{System Model and Evaluation Metrics}\label{sTwo}
In this section, we will introduce the system model and corresponding evaluation metrics.

\subsection{System Model}
To address the frequency leakage problems, We want the adversarial energy to be concentrated in the spectrum where the signal is located.
We first formalize the mathematical model of the frequency adversarial attack method with time and frequency domain Concealment.

\begin{equation}
\arg \min _{\delta} \frac{\operatorname{energy}\left(\delta_{\text {so }}\right)}{\operatorname{energy}(\delta)} \text { s.t. } f(x+\delta) \neq y \text { and }\|\delta\|_{\infty} \leq \zeta,
\end{equation}
where, $\operatorname{energy}\left(\cdot\right)$ is the energy calculation function;
$\delta$ is the adversarial perturbation;
$\delta_{\text {so }}$ is the part outside the focus spectrum;
$f\left(\cdot\right)$ is the AMC model;
$x$ is the original signal example and $y$ is the correct label;
$\zeta$ is the maximum allowable infinity norm.
This paper hopes to obtain the perturbation that can fool the AMC model.
The infinity norm of this perturbation is within a certain range, and the proportion of power outside the focus spectrum is as small as possible.

\subsection{Adversarial Evaluation Metrics}
In order to conduct a comprehensive analysis, we design the perturbation out-spectrum energy ratio (OSER) to measure the frequency domain concealment and use the fitting difference (FD) ~\cite{9322088} to evaluate the signal time domain quality.

\subsubsection{Out-Spectrum Energy Ratio}
The out-spectrum energy of adversarial perturbations makes adversarial examples easier to detect.
Therefore, we design an indicator as \eqref{eqOSER} to measure the energy ratio of the adversarial perturbation out-spectrum of the original signal relative to the total perturbation.


\begin{equation}
\textit{OSER}=10 \log \left(\frac{\sum_{\mathrm{B} \cdot \mathrm{L} \mathrm{f}_{s}<i<L-\mathrm{B} \cdot \mathrm{L} \mathrm{f}_{\mathrm{s}}}^{L} s^{2}(i)}{\sum_{i}^{L} s^{2}(i)}\right).
\label{eqOSER}
\end{equation}

\subsubsection{Fitting Difference}
The infinity norm can only reflect the maximum difference of a sample point in a segment of the signal, but not the overall quality.
Therefore, we use the FD \cite{9322088} as an evaluation Indicator for the signal quality in the time domain:

\begin{equation}
\textit{FD}=\frac{\sum_{i=1}^{L}\left(x_{i}-x_{i}^{\prime}\right)^{2}}{\sum_{i=1}^{L}\left(x_{i}-\bar{x}\right)^{2}}.
\label{eqFD}
\end{equation}

\section{Adversarial Attacks Algorithm}\label{sThree}
In this section, we will introduce the proposed SFFAA algorithm, as well as Meta-SFFAA in black-box attack.

\subsection{Spectrum Focused Frequency Adversarial Attacks}

\begin{figure}[htbp]
\centerline{\includegraphics[width=0.9\linewidth]{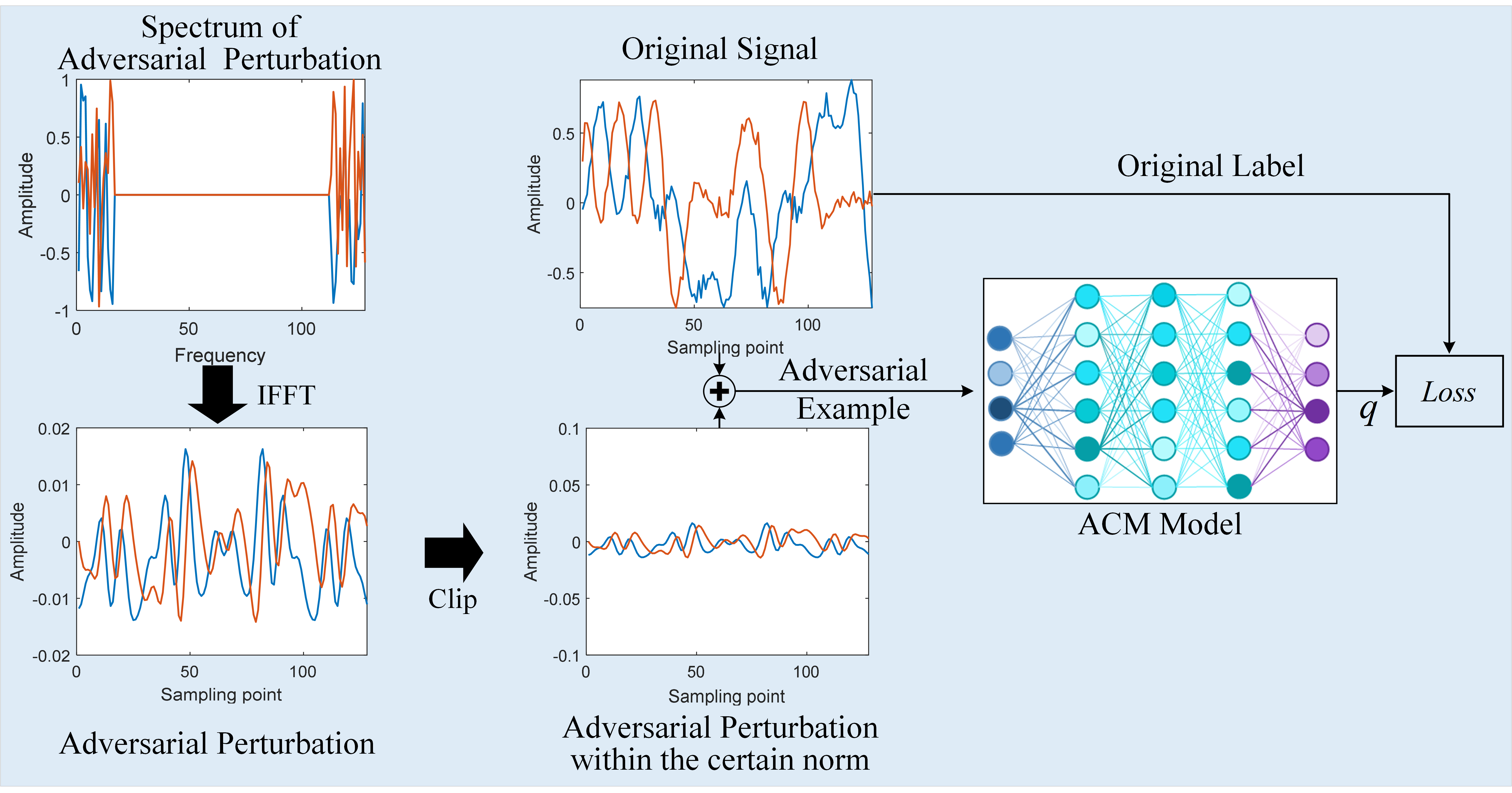}}
\caption{Overview of the proposed SFFAA algorithm.}
\label{OverviewSFFAA}
\end{figure}

Based on the above modeling and analysis of adversarial attack problem, we design and propose SFFAA algorithm.
The algorithm starts from the frequency domain, which ensures that the adversarial disturbance is focused in the spectrum.
Further, the truncation function will control the infinity norm of the perturbation.
Fig.~\ref{OverviewSFFAA} presents the overview of the proposed SFFAA algorithm.

First, initialize the random spectrum data in focus location, and use the Inverse Fast Fourier Transform (IFFT) to get the corresponding time domain adversarial perturbations.
In order to ensure that the infinite norm is within a certain range, we truncate the adversarial perturbations with $\operatorname{Clip}(\cdot)$ operation as \eqref{eqClipIfft}.
Note that the $\operatorname{Clip}(\cdot)$ function may introduce high frequency components, but these components are very small.

\begin{equation}
\delta_{0}=\operatorname{Clip}\left(\operatorname{IFFT}\left(s_{0}\right)\right),
\label{eqClipIfft}
\end{equation}
where, $s_{0}$ is initialized spectrum data.
Half its number of sample points can be calculated by $B\cdot L / f_{s}$;
$B$ is the set focus bandwidth, which is generally the signal bandwidth.
$L$ is the length of the original signal.
$f_{s}$ is the sampling rate of the original signal;
$\delta_{0}$ is the adversarial perturbation corresponding to $s_{0}$;
Furthermore, the adversarial perturbation is entered into the AMC model with the original signal, and confidence list $q$ is obtained.
A loss function is designed between the correct label $l$ and $q$ to punish the corresponding confidence as \eqref{eqLoss}.

\begin{equation}
L(q, l)=-\log \left(1-q_{l}+\varepsilon\right),
\label{eqLoss}
\end{equation}
where, $q_{l}$ is the confidence corresponding to the correct label.
$\varepsilon=1e-6$ is used to prevent $1-q_{l}$ being 0.
The update rule for $s$ are given in \eqref{eqUpdate}.

\begin{equation}
s_{n+1}=s_{n}+\alpha \cdot \operatorname{sign}\left(\nabla_{s_{n}} L(q, l)\right),
\label{eqUpdate}
\end{equation}
where, $\alpha$ is the update step size;
$n$ is the current number of updates;
Using \eqref{eqClipIfft} and \eqref{eqUpdate} to iteratively update s0 for $N$ times, the final perturbation spectrum and adversarial perturbation can be obtained.
The pseudo code is shown in Algorithm 1.

\begin{algorithm}[h]
\caption{SFFAA}
\label{alg:1}
\small
\begin{algorithmic}[1]
\renewcommand{\algorithmicrequire}{\textbf{Input:}}
\renewcommand{\algorithmicensure}{\textbf{Output:}}
\REQUIRE {The AMC model $f(\cdot)$;
The initialized spectrum data $s_{0}$;
The original data $x$;
The correct label $l$;
The update step size $\alpha$.

\ENSURE {The final perturbation spectrum $s_{n}$;
The final adversarial example $x^{\prime}$.}\\

\STATE {Randomly initialize $s_{0}$;}
\FOR{epoch $n$ = 0 to $N -$ 1 }
  \STATE{Take the IFFT for $s_{n}$ and use the truncation function to get the adversarial perturbation $\delta_{n}$;}
  \STATE{Superimpose adversarial perturbation $\delta_{n}$ to get adversarial example $x_{n}^{\prime}$;}
  \STATE{Feed the adversarial example $x_{n}^{\prime}$ into the model to get confidence list $q$;}
  \STATE{According to the loss function \eqref{eqLoss}, calculate the loss value between $q$ and $l$;}
  \STATE{Get the gradient of $s_{n}$ by backpropagation;}
  \STATE{Get the $s_{n+1}$ by \eqref{eqUpdate} according to $\alpha$;}
  \ENDFOR
\RETURN {$s_{N}$ and $x_{n}^{\prime}$;}
}
\end{algorithmic}
\end{algorithm}

\subsection{Meta Spectrum Focused Frequency Adversarial Attacks}
As with adversarial attack algorithm in time domain, the adversarial samples generated by the SFFAA algorithm under the white-box attack do not have strong black-box transferability.
It is difficult to obtain the ideal attack effect by directly applying the adversarial samples in the white box attack scenario to a black-box one.
It is due to differences in model decision boundaries with differences in model initialization, connection structure, training optimizers, etc.

\begin{figure*}[htbp]
\centerline{\includegraphics[width=0.8\linewidth]{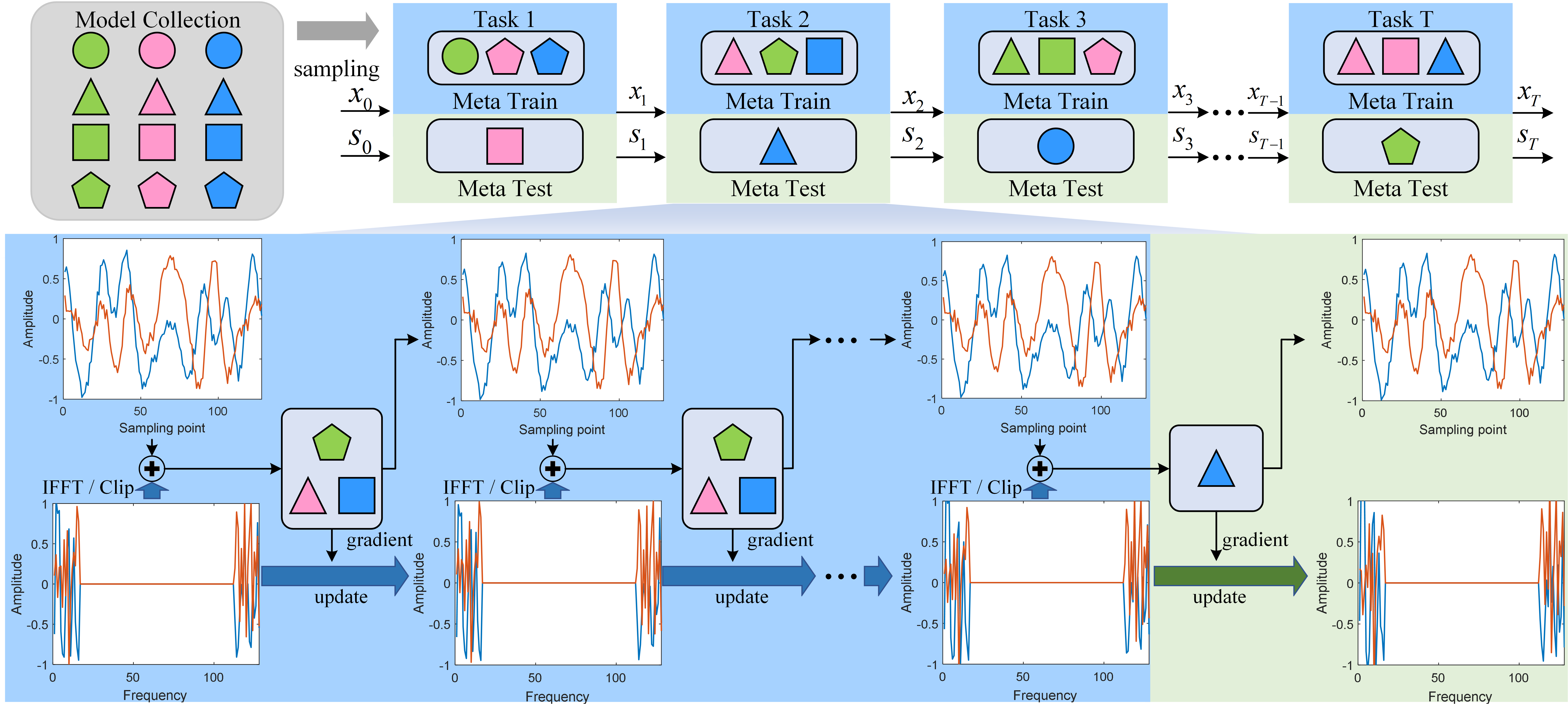}}
\caption{Overview of the proposed Meta-SFFAA algorithm.}
\label{OverviewMSFFAA}
\end{figure*}

To this, we draw on the concept of meta gradient adversarial attack~\cite{Yuan_2021_ICCV}, and propose a Meta-SFFAA algorithm, as shown in Fig.~\ref{OverviewMSFFAA}.
The algorithm consists of multiple tasks including meta-train and meta-test.
At each task, meta-train is used to simulate the white-box attack, while meta-test is used to simulate the black-box attack.
Each task is performed sequentially, so that meta-train and meta-test are run alternately.
The gradient difference between the black-box and the white-box is adaptively reduced, thereby enhancing transferability in the black-box attack.

Specifically, we first build a Model Collection containing $M$ different models.
Note that the Model Collection should not contain the model for testing black-box attack.
Then, $R+1$ models are randomly selected from the Model Collection to form a task, and a total of T tasks are generated.
In each task, R models are used for meta-train and the other one is used for meta-test.

During the meta-train process, the average logistic output of the R models is computed as the final output.
The final confidence list $q_{t,k}$ can be obtained by further $\operatorname{softmax}(\cdot)$ function as:

\begin{equation}
q_{t, k}=\operatorname{softmax}\left(\sum_{r} \operatorname{logistic}\left(r, x_{t, k}^{\prime}\right) / R\right),
\label{eqQTK}
\end{equation}
where, $x_{t, k}^{\prime}$ is the input data for the $k$-th meta-train in the $t$-th task;
$\operatorname{logistic}(r, \cdot)$ is the output logistic of the $r$-th model.
Then, One-step update is done by computing the gradient using the loss function as shown in \eqref{eqSt}.

\begin{equation}
s_{t, k+1}=s_{t, k}+\alpha \cdot \operatorname{sign}\left(\nabla_{s_{t, k}} L\left(q_{t, k}, l\right)\right).
\label{eqSt}
\end{equation}

After performing meta-train for $K$ times, perform meta-test for one time.
The meta-test is the same as the original algorithm, the only difference is that the model is a random one.
Therefore, the meta-testing process will not be repeated here.
The adversarial example $x_{t}$ and adversarial perturbation spectrum $s_{t}$ generated by each task are passed on to the next task until all tasks are completed.
The final $x_{T}$ is an adversarial example with strong black-box transferability.

\section{Experimental Evaluation}\label{sFour}
In this section, We conduct extensive experiments to qualitatively and quantitatively analyze the performance of the proposed algorithm.

\subsection{Datasets}
The generation scheme of the dataset, and the constructed baseline network structure are introduced as the basis for subsequent experiments.
We used MATLAB to generate simulation datasets including eight modulation types, BPSK, QPSK, 8PSK, 16QAM, 64QAM, PAM4, GFSK, and 2FSK.
The channel model is a Rician channel with additive white Gaussian noise.
The dataset parameters are shown in Table \ref{tbDP}.

\begin{table}[htbp]
\caption{Datasets Parameters.}
    \centering
    \begin{tabular}{|c|c|}
    \hline
        Item & Value \\ \hline
        Symbol rate & 1MHz \\ \hline
        Baseband sampling rate & 8MHz \\ \hline
        Carrier frequency & 300Mhz \\ \hline
        Direct path frequency offset & 2Hz \\ \hline
        Indirect path time delay & 0.3us \\ \hline
        Indirect path maximum frequency shift & 2Hz \\ \hline
        Average path gain & -5.9dB \\ \hline
        Rician factor & 10 \\ \hline
        Signal-to-Noise Ratio (SNR) & -20dB: 2dB: 18dB \\ \hline
        Example size & 2×128 \\ \hline
        Example number & 1,000/type/SNR \\ \hline
        Train: Valid: Test & 8:1:1 \\ \hline
    \end{tabular}
\label{tbDP}
\end{table}

\subsection{Comparison of White-box Attack Performance}
\begin{figure}[htbp]
\centerline{\includegraphics[width=1.0\linewidth]{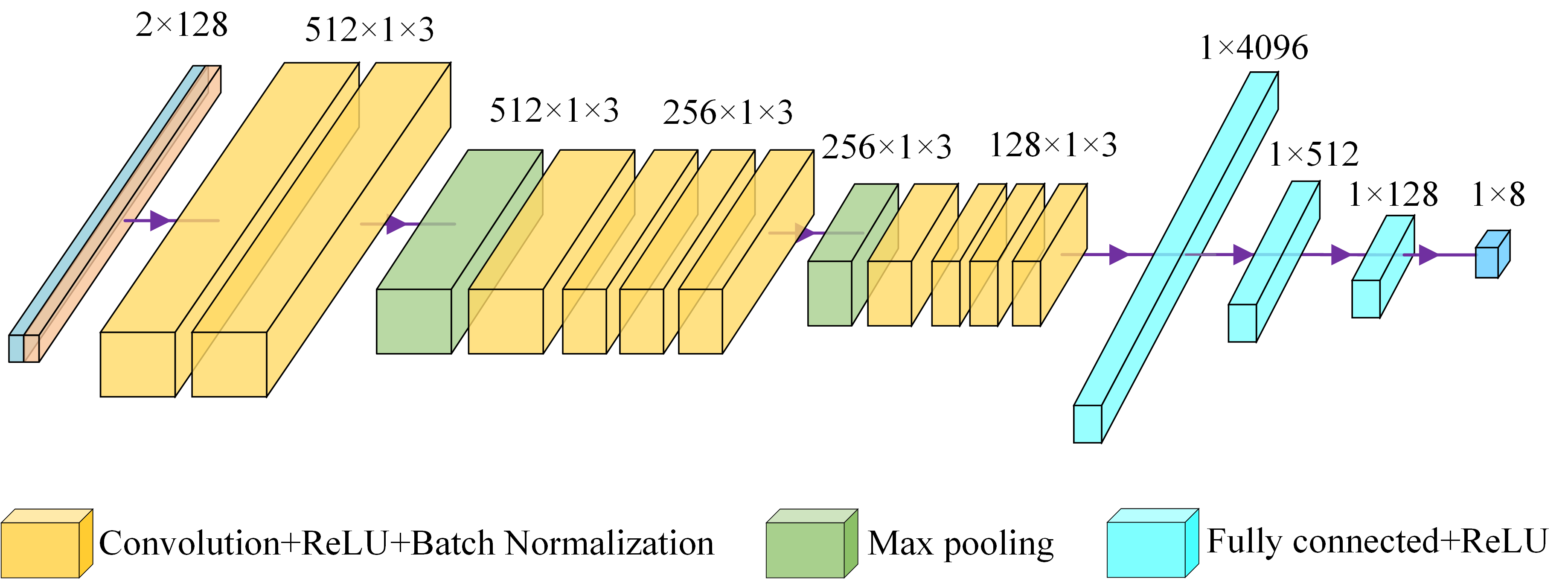}}
\caption{Architecture diagram of the baseline model.}
\label{Baseline}
\end{figure}

White-box attacks are the most ideal and direct indicators to evaluate the attack performance of an attacking algorithm.
We built a custom model as the baseline model for this experiment, as shown in Fig.~\ref{Baseline}.
The Baseline model is used as the target model.
Before comparing the proposed SFFAA algorithm with other algorithms, we first conduct an exploration experiment on the number of iteration steps.
The perturbation strengths is set to 0.1, the iteration step size is set to 0.05, and the number of iteration steps is set to 20, 40, 60 and 80, respectively.
According to the experimental results, in order to obtain better attack performance and avoid serious overfitting, we set the number of iteration steps of the SFFAA algorithm to 50 in the following experiments.

\begin{figure}[htbp]
\centerline{\includegraphics[width=0.9\linewidth]{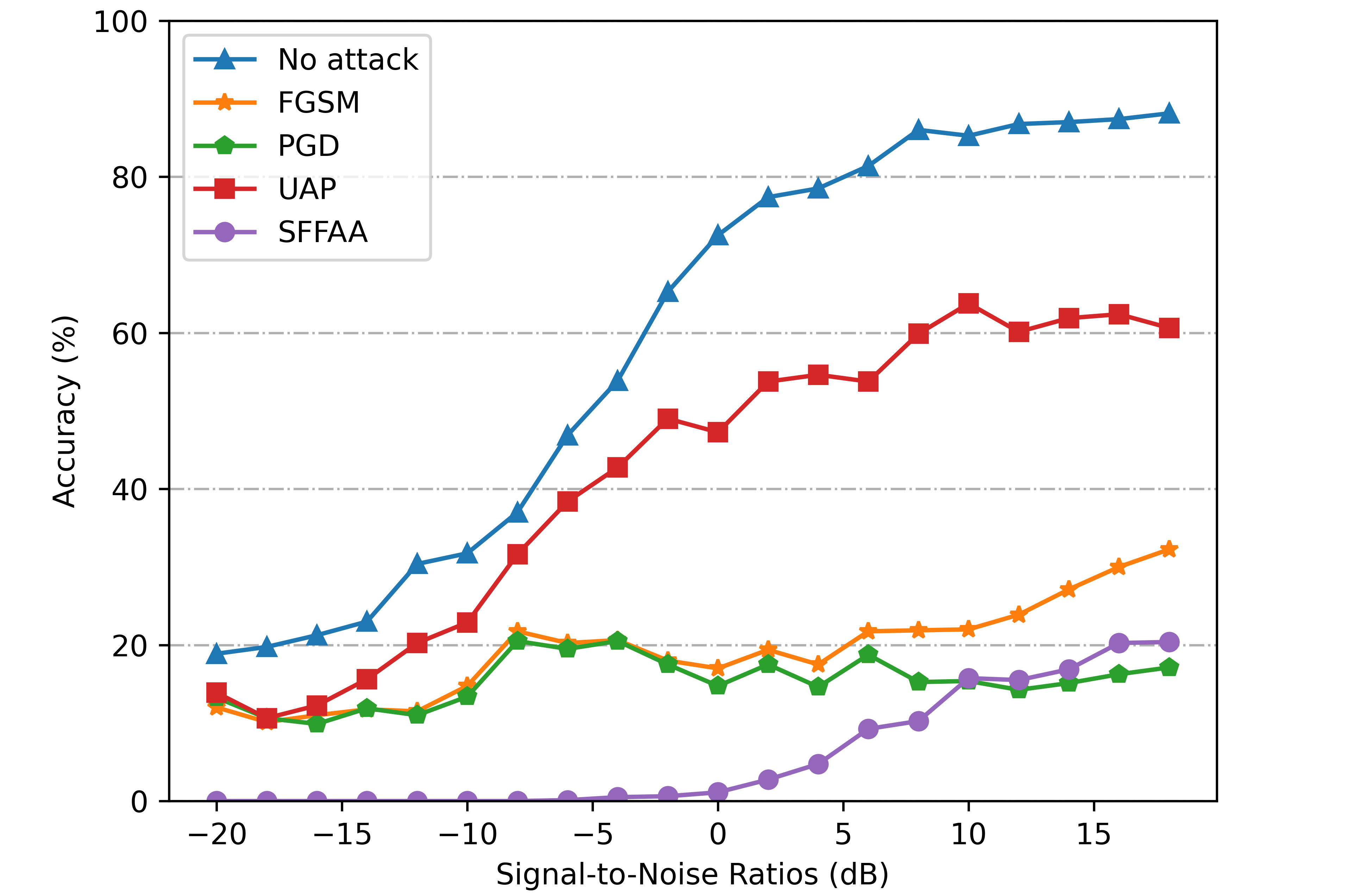}}
\caption{Comparison of attack performance in white-box attacks.}
\label{whiteFour}
\end{figure}

For fully evaluate the attack performance of the proposed SFFAA algorithm, the fast gradient sign method (FGSM)~\cite{2014arXiv1412.6572G}, projected gradient descent (PGD)~\cite{2017arXiv170606083M} and universal adversarial perturbations (UAP)~\cite{Moosavi-Dezfooli_2017_CVPR} algorithms are used for comparison.
In the experimental setup, the perturbation strengths of the four algorithms are all 0.1, and the iteration step size and the number of iteration steps of the PGD and UAP algorithm are 0.02 and 10, respectively.
The classification accuracy under no attack and the four attack algorithms is shown in Fig.~\ref{whiteFour}.

As can be seen from the Fig.~\ref{whiteFour}, the four attack algorithms cause the accuracy to decrease to varying degrees.
The UAP algorithm generates one adversarial perturbation for the entire datasets, rather than design for each example, so its attack is relatively weak.
The iterative algorithm can be updated many times to find a better path to reduce the target confidence, so the attack performance of the PGD is stronger than that of the single-step attack FGSM.
By contrast, SFFAA can completely fool the target model at low SNR.
As the SNR increases, the attack performance weakens.
When the SNR is higher than 10dB, the attack performance is slightly weaker than that of the PGD.
Compared with the above three attack algorithms, the SFFAA algorithm has the strongest attack performance.

\subsection{Comparison of Black-box Attack Performance}
\begin{figure}[htbp]
\centerline{\includegraphics[width=0.9\linewidth]{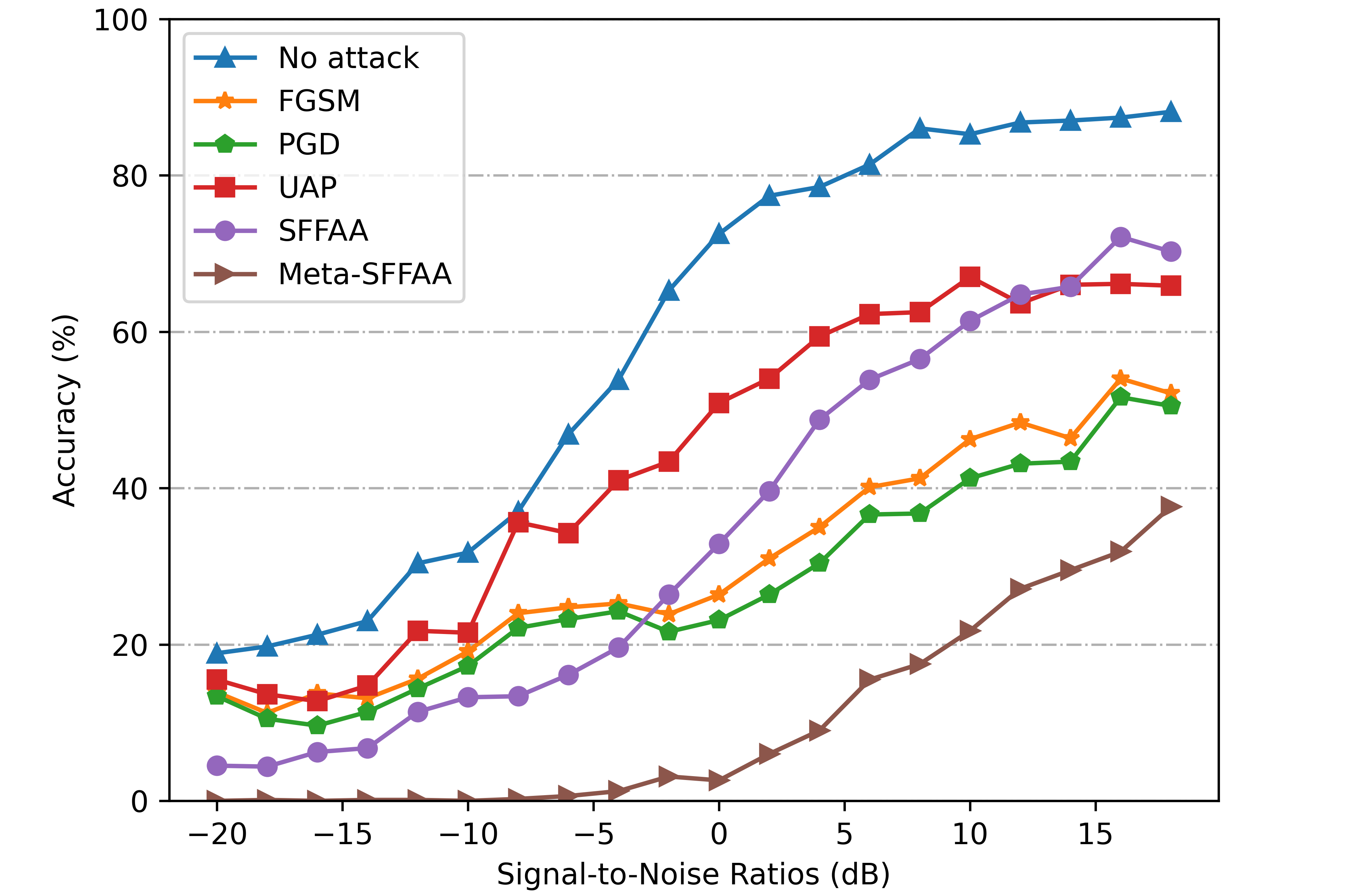}}
\caption{Comparison of attack performance in black-box attacks.}
\label{blackFive}
\end{figure}

Compared with the white-box attack, the black-box attack is closer to the real scenario, that is, the attacker has no information of the structure and training method of the target model.
Next, we compare the black-box attack performance of the SFFAA algorithm with FGSM, PGD, and UAP.
Further, we propose to learn from meta-learning to improve the black-box attack transferability.
In the Meta-SFFAA algorithm, we use three kinds of optimization algorithms, RMSprop, Adam, and Adamax, respectively to train four kinds of models, baseline, ResNet18, ResNet50, and VGG11, and get the model collection. 
The black-box target model is GoogLeNet.
4 models were randomly selected for each task, three of which are used for meta-train for 50 times and one is used for meta-test for one time.
15 tasks like this are created.
The classification accuracy of the black-box target model under these attack algorithms is shown in Fig.~\ref{blackFive}.

Comparing Fig.~\ref{whiteFour} and Fig.~\ref{blackFive}, it can be seen that the attack performance of the four algorithms in the black-box attack is weakened.
Among them, the performance of the UAP algorithm is reduced the least, because the transferability of the adversarial examples is fully considered in its design.
The attack performance of FGSM and PGD have similar weakening degree, but PGD is still better than the FGSM algorithm.
The most weakened attack performance is the proposed SFFAA algorithm.
Obtain very impressive attack performance in white-box attack, which tends to be overfit, so it is reasonable to have maximum decay under black-box attack.
Finally, the attack performance of the Meta-SFFAA is greatly improved compared to the SFFAA algorithm.
Meta-learning integrates the gradient directions of multiple models and simulates the process of alternating white-box and black-box, enabling SFFAA to obtain excellent transferability to black-box attacks.

\subsection{Comparison of Time and Frequency Domain Performance}
After comparing the performance of white-box and black-box attacks, we further conduct qualitative and quantitative analysis of the proposed SFFAA algorithm in the time and frequency domains.

\subsubsection{Perturbation Spectrum Distribution}%
\begin{figure}[htbp]
\centering
\scriptsize
\subfigure[\scriptsize FGSM]{\includegraphics[width=0.48\linewidth]{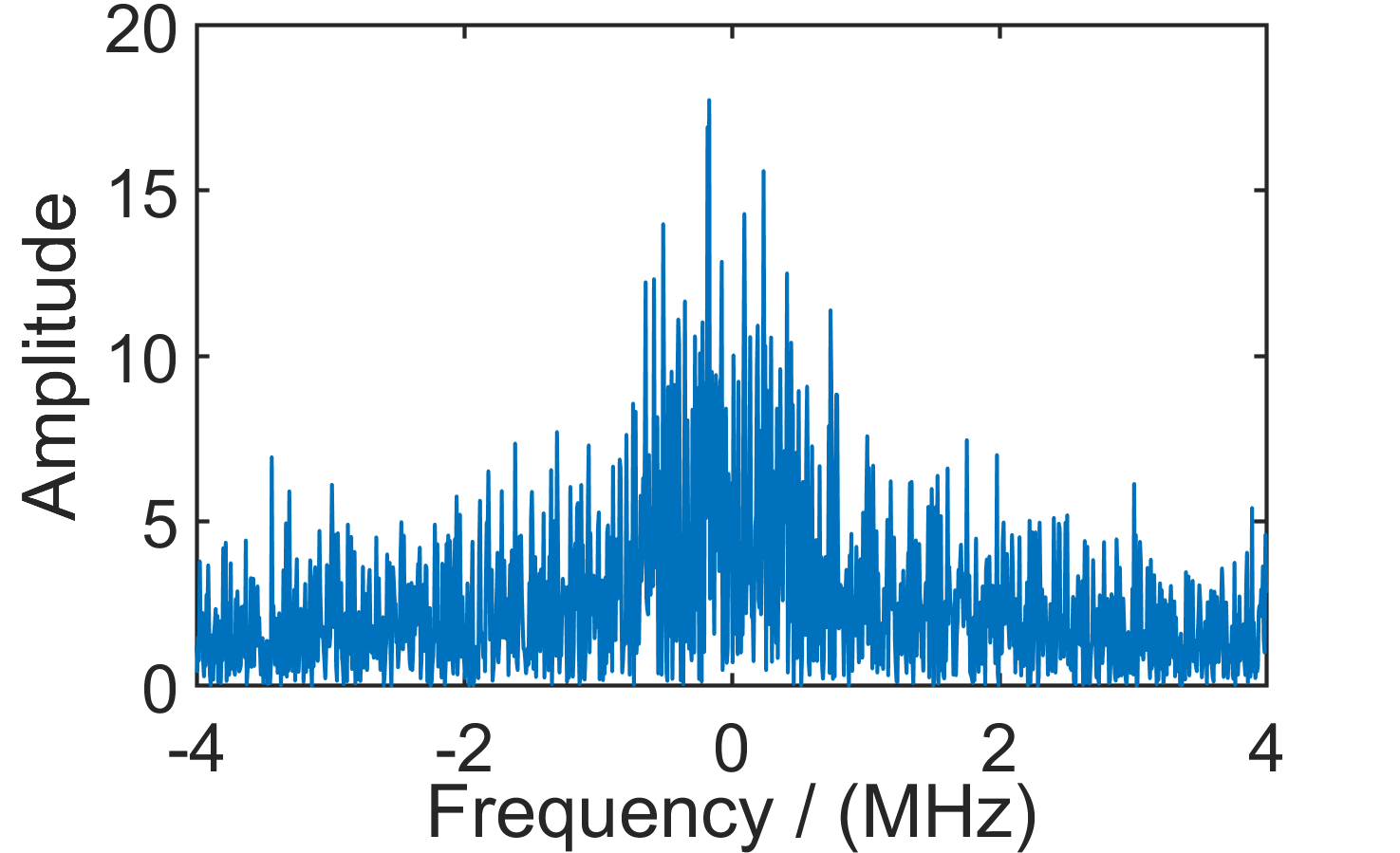}
\label{pSpecDFGSM}}
\hfill
\subfigure[\scriptsize PGD]{\includegraphics[width=0.48\linewidth]{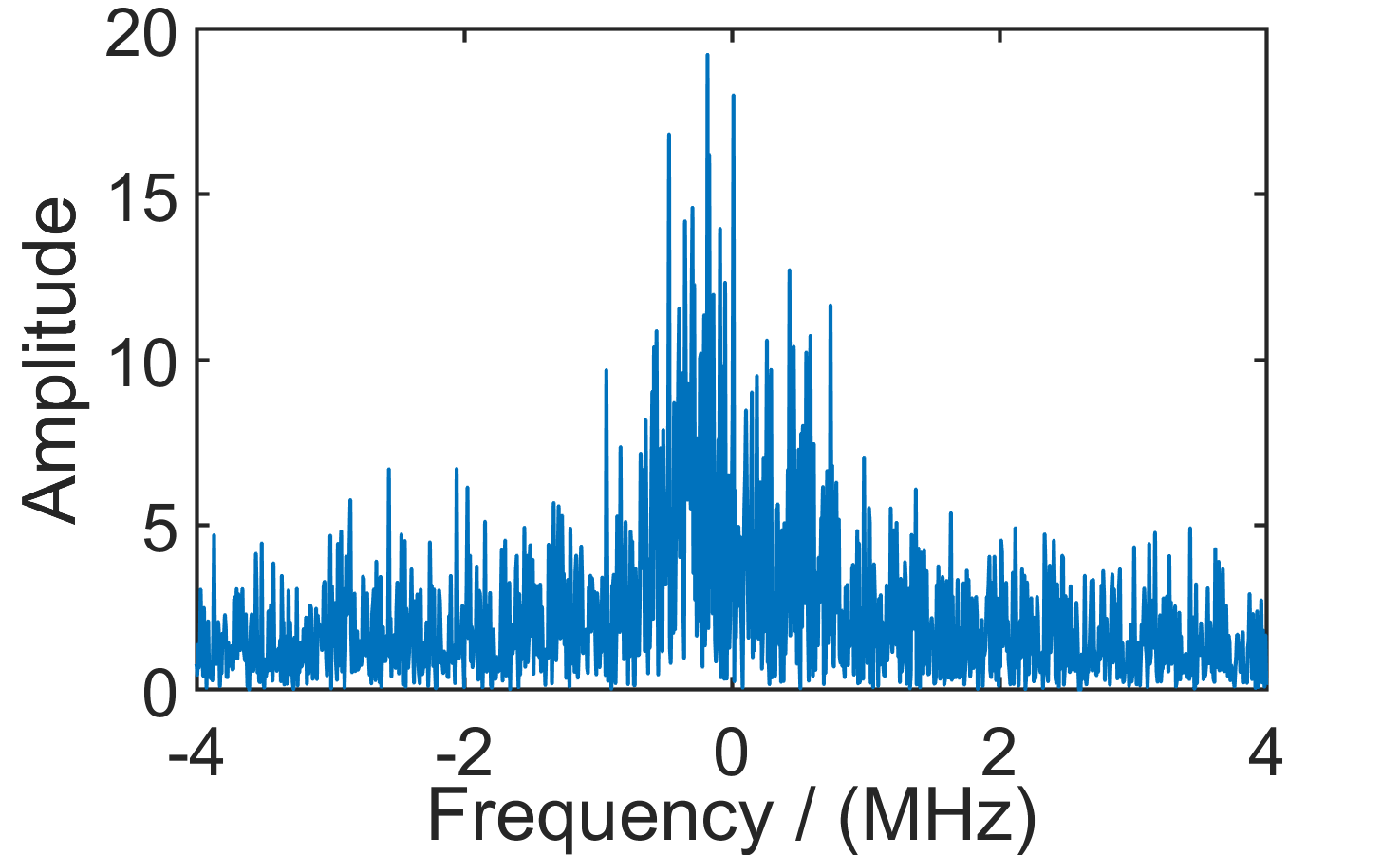}
\label{pSpecDPGD}}
\hfill
\subfigure[\scriptsize UAP]{\includegraphics[width=0.48\linewidth]{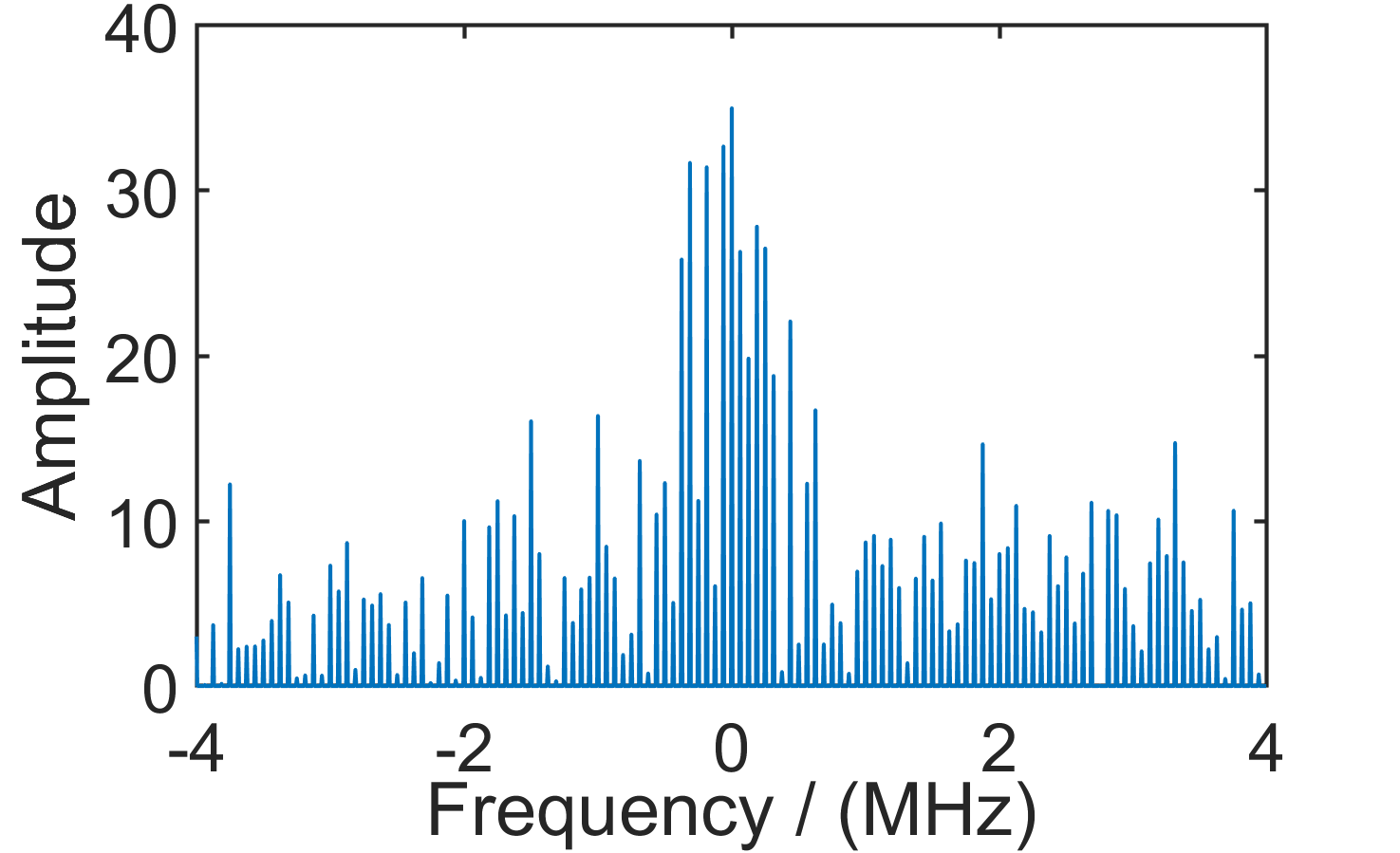}
\label{pSpecDUAP}}
\hfill
\subfigure[\scriptsize SFFAA]{\includegraphics[width=0.48\linewidth]{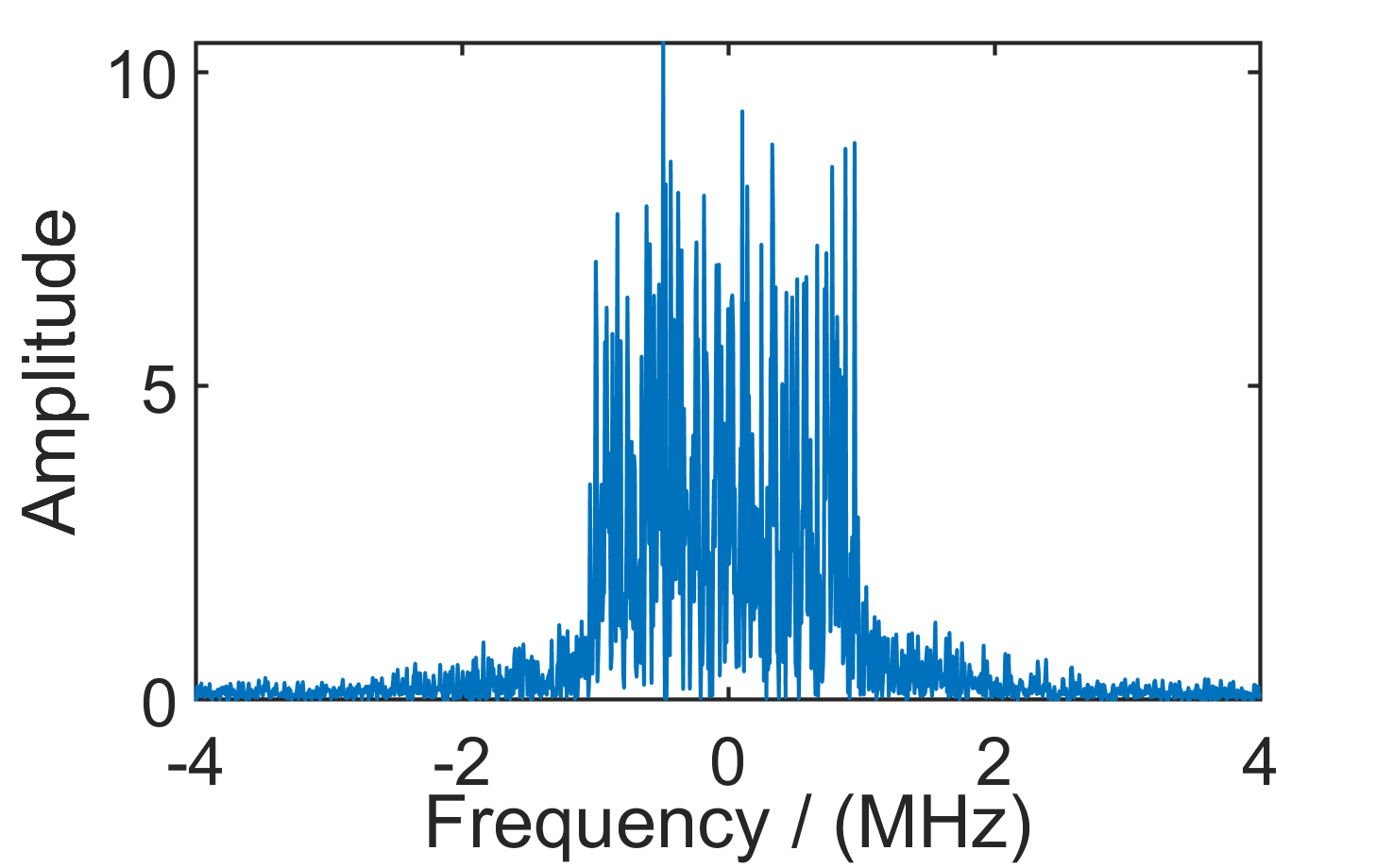}
\label{pSpecDSFFAA}}
\caption{Perturbation spectrum distribution for different algorithms.}
\label{pSpecD}
\end{figure}

A qualitative analysis method in the frequency domain is to plot the perturbation spectrum distribution to assess the concentration of the perturbation in the spectrum.
We plot the spectrum distribution of the perturbations of the four algorithms as shown in Fig.~\ref{pSpecD}.

As can be seen from the Fig.~\ref{pSpecD}, FGSM, PGD and UAP algorithms have a lot of energy outside the frequency band (1MHz) where the signal is located.
Even, the UAP algorithm produces many spikes in the sidebands.
The energy of the sidebands, as well as these spikes, reduces the concealment of the adversarial signal.
In contrast, the SFFAA algorithm has an obvious focus, which concentrates the attack energy in the spectrum focus, thus obtaining a significant attack performance and excellent concealment.

\subsubsection{Fitting Difference}
\begin{figure}[htbp]
\centerline{\includegraphics[width=0.9\linewidth]{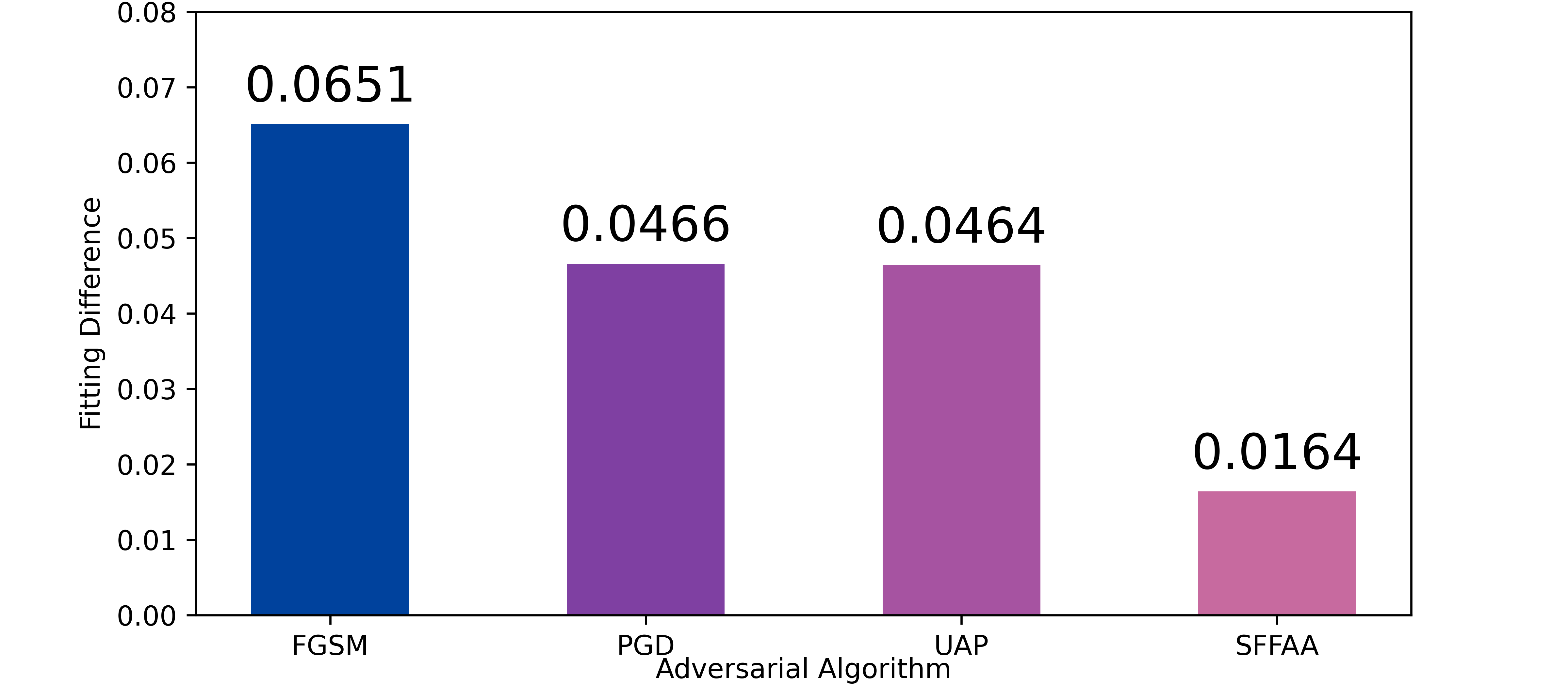}}
\caption{Fitting difference for different algorithms.}
\label{FD}
\end{figure}

The FD reflects the magnitude of the difference between the adversarial example and the original signal.
We randomly select 80 signals from the datasets and generate adversarial examples using the four algorithms.
The average FD for these examples were calculated and shown in Fig.~\ref{FD}.

It can be found that the FD of the FGSM algorithm is the largest,
and the PGD and UAP algorithms are similar and second only to the FGSM.
The FD of the SFFAA algorithm is the smallest, which is close to a quarter of that of the FGSM algorithm.
That is because the perturbation of the SFFAA is concentrated in the low frequency region and is more flexible, so there is minimal waste of extra.

\subsubsection{Out-Spectrum Energy Ratio}%
\begin{figure}[htbp]
\centerline{\includegraphics[width=0.9\linewidth]{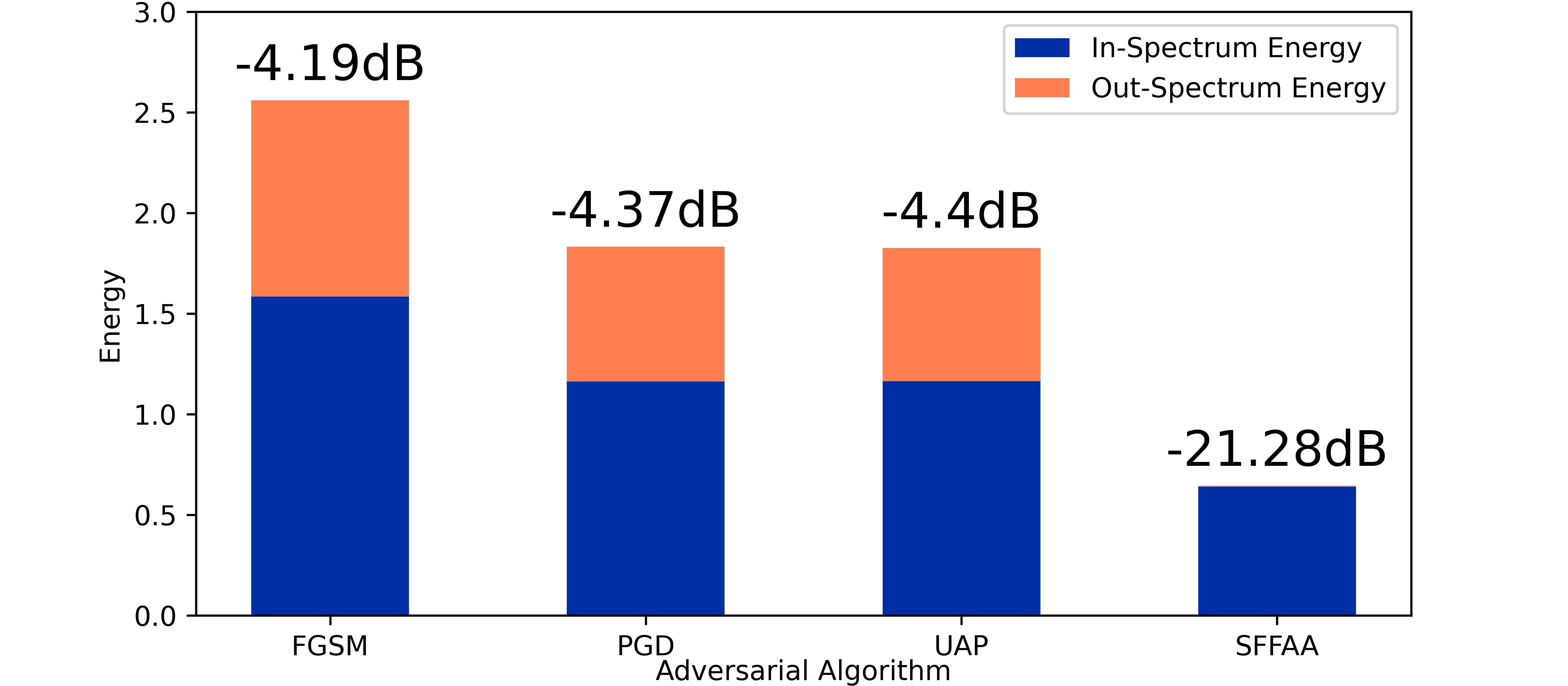}}
\caption{Out-spectrum energy ratio for different algorithms.}
\label{OSER}
\end{figure}

In order to quantitatively analyze the spectral concentration of the adversarial attack algorithm in the frequency domain,
we calculate the average in-spectrum and out-spectrum energies and OSER for the perturbations of the 80 examples, as shown in Fig.~\ref{OSER}.

The energy for each algorithm consists of the in-spectrum and out-spectrum energies, and OSER is marked above the energy column.
Among them, the energy of the perturbation of the FGSM algorithm is the largest, while the energy of the PGD and UAP algorithms is at a medium level, and the energy of the SFFAA algorithm is the smallest.
Also, the OSER of FGSM, PGD and UAP algorithms are very close.
Out-spectral energy is between one-third and one-quarter of the total energy.
However, the out-spectrum energy of the SFFAA algorithm is only one-hundredth of the total energy.
This proves that the SFFAA can concentrate the attack energy on the spectrum focus.

\section{Conclusion}\label{sFive}
Adversarial attacks and defenses against AI-based AMC models are critical topics to promote efficient, secure and credible applications of AI in electromagnetic space.
We found the problem that the existing attack algorithms will introduce more high-frequency components in the frequency domain,
due to they are updated bluntly in the time domain.
Such adversarial attacks can be easily detected.
For this issue, we design a SFFAA method from the frequency domain.
In addition, we also draw on the idea of meta-learning and design the Meta-SFFAA algorithm to improve the transferability in black-box attacks.
Qualitative indicators including perturbation spectrum distribution, and quantitative indicators including fitting difference, and out-spectrum energy ratio are calculated to demonstrate that the proposed algorithm can concentrate the adversarial energy in the spectrum focus,
to obtain higher attack performance.
At the same time, the concealment of the spectrum is maintained.

However, the proposed algorithm still has shortcomings.
The time complexity of the algorithm is relatively high, which will affect the transmission rate of the adversarial signal.
Besides, the research is in the direct-access adversarial attack scenario.
In the future, we will study the efficient generation method, and will also develop a spectrum focused frequency adversarial attack algorithm at the transmitter to enhance the applicability in practical scenarios.
Undoubtedly, this paper provides a new perspective for the adversarial attack in the electromagnetic field.
This perspective considers more characteristics of the electromagnetic field and has important research value.

\section*{Acknowledgment}
This work is supported by the Fundamental Research Funds for the Central Universities (3072022CF0804, 3072022CF0601).
This work is also supported by the Key Laboratory of Advanced Marine Communication and Information Technology, the Ministry of Industry and Information Technology, and Harbin Engineering University, Harbin, China.

\bibliographystyle{IEEEtran}
\bibliography{../../cite_SC}

\vspace{12pt}

\end{document}